\documentclass[aps,prl,twocolumn,showpacs,floatfix,superscriptaddress]{revtex4}
\usepackage{graphicx,subfigure}
\usepackage{amsmath,amssymb}
\usepackage{float}
\usepackage[hidelinks]{hyperref}
\usepackage{xcolor}
\begin{document}
\title{Towards a universal  description of hadronic phase of QCD  }
\author{Aman Abhishek}
\email{amanabhi@imsc.res.in}
\affiliation{The Institute of Mathematical Sciences, a CI of 
Homi Bhabha National Institute, Chennai, 600113, India}
\author{Sayantan Sharma}
\email{sayantans@imsc.res.in}
\affiliation{The Institute of Mathematical Sciences, a CI of 
Homi Bhabha National Institute, Chennai, 600113, India}
%
%
\begin{abstract}
Mean-field model quantum field theories of hadrons were traditionally developed to describe 
cold and dense nuclear matter and are by now very well constrained 
from the recent neutron star merger observations. We show that when 
augmented with additional known  hadrons and resonances 
but not included earlier, these mean-field models can be extended beyond 
its regime of applicability. Calculating some specific ratios of baryon 
number susceptibilities for finite temperature and moderate values of 
baryon densities within mean-field approximation, we show that these match 
consistently with the lattice QCD data available at lower densities, unlike 
the results obtained from a non-interacting hadron resonance gas model. We also estimate the 
curvature of the line of constant energy density, fixed at its corresponding value 
at the chiral crossover transition in QCD, in the temperature-density plane. The 
number density at low temperatures and high density is found to be about twice the 
nuclear saturation density along the line of constant energy density of $\epsilon=348
\pm 41$ MeV/$\text{fm}^3$. Moreover from this line we can indirectly constrain the 
critical end-point of QCD to be beyond $\mu_B=596$ MeV for temperature $\sim 125 $ MeV.
\end{abstract}

\maketitle

\textbf{Introduction}
Developing an effective field theory description of hadrons preceded the 
discovery of the field theory of strong interactions, Quantum Chromodynamics 
(QCD). Indeed, based on the observation of the exponential increase in 
the density of states of hadrons with increasing temperature, it was proposed 
that hadronic matter will undergo a phase transition to a deconfined phase~\cite{Hagedorn:1965st}. 
Ab-initio lattice studies have confirmed this scenario and showed the 
existence of a smooth crossover at zero baryon 
density~\cite{Aoki:2006we,Bazavov:2011nk,Bhattacharya:2014ara,Burger:2018fvb,Taniguchi:2020mgg} 
from a hadron phase to a quark-gluon plasma phase in $2+1$ flavor QCD  with physical 
quark masses at a temperature $T_c=156.5\pm 1.5$ MeV~\cite{HotQCD:2018pds}. 
Furthermore lattice QCD techniques have now provided us with 
the state-of-the art Equation of State (EoS) of hadrons in the continuum 
limit~\cite{Borsanyi:2013bia,HotQCD:2014kol,Ding:2015fca}. 
Such reliable results have boosted the efforts for understanding the different hadron 
interactions and develop effective relativistic quantum field theories of hadrons,  
the so-called \emph{hadrodynamics}. Constraining \emph{hadrodynamics} to a very good extent 
is of fundamental importance in understanding QCD at finite temperature and density.

A description of the hadron phase in terms of a gas of non-interacting hadrons and 
the narrow-width resonances (HRG)~\cite{Dashen:1969ep,Dashen:1974jw,Dashen:1974yy} 
has been shown to describe bulk thermodynamic observables in QCD, e.g. free-energy
~\cite{Karsch:2003vd,Karsch:2003zq,Huovinen:2009yb} and chiral 
condensate~\cite{Borsanyi:2010bp,Biswas:2022vat} to a surprisingly good accuracy. 
This description is the basis for statistical hadronization models that have been very 
successful in describing the experimental yields of different hadron species in 
heavy-ion colliders~\cite{Andronic:2017pug}. A justification of this comparison 
came from the observation that non-resonant part of the phase shifts of the 
attractive hadron interactions largely cancel out in the calculation of free-energy, 
and the interacting part of the pressure can be well described by the contribution 
of resonances treated as stable particles~\cite{Venugopalan:1992hy}. However 
with increasing precision of the lattice data on fluctuations of conserved numbers like 
baryon number, strangeness and electric-charge, a visible departure from the HRG 
model predictions are by now clearly evident. Extension of the basic HRG model by 
augmenting it with the many not-yet experimentally measured baryon resonances
~\cite{Majumder:2010ik} mainly in the strangeness sector~\cite{Bazavov:2013dta}, but 
predicted from lattice QCD and different relativistic quark models, termed as QMHRG 
can explain many puzzles like simultaneous freezeout of light, strange and open-charm  
hadrons~\cite{Bazavov:2014xya,Bazavov:2014yba,Mukherjee:2015mxc}.  However there are 
thermodynamic observables in QCD which cannot be yet explained within the QMHRG model, 
specially close to $T_c$~\cite{Andronic:2017pug,Bollweg:2021vqf,Karthein:2021cmb}, 
highlighting the importance of non-resonant and repulsive interaction channels.

Repulsive interactions between baryons will become more important 
at large baryon densities. However constraining them is challenging~\cite{Lovato:2022vgq} 
as there are no first principles calculations of the EoS available yet from 
lattice QCD due to the infamous \emph{sign-problem}~\cite{deForcrand:2009zkb,Schmidt:2017bjt}. 
Experimental constraints are also few and come mainly 
from the study of supernovae, neutron star mergers and nuclear matter from 
the low energy heavy-ion collision experiments at CERN SPS and HADES, Darmstadt 
and in future from the upcoming FAIR facility at GSI Darmstadt and NICA at Dubna. Recent 
advances in the multi-messenger astronomy of neutron stars have opened 
a new avenue to constrain the nuclear models and its 
EoS~\cite{Annala:2021gom,Lovato:2022vgq,Sumiyoshi:2022uoj}.  
In this regime of high baryon densities there are a multitude of nuclear models with 
different EoS. These are usually based on the 
Dirac-Brueckner-Hartree-Fock~\cite{Brockmann:1984qg, TerHaar:1986xpv, deJong:1997mn}
approach or relativistic mean field models~\cite{Serot:1984ey, Serot:1997xg, Bodmer:1991hz}. 
In the former approach parameters of the interactions are fixed from 
experimental inputs of nucleon-nucleon and nucleon-meson scatterings. However its 
application to finite density nuclear matter remains difficult. On the other hand 
in relativistic hadron models, pioneered by Walecka, the interactions between the 
nucleons are implemented at mean-field level by coupling to effective meson degrees 
of freedom. The parameters of the interaction terms are instead determined by matching 
to the empirical saturation properties of nuclear matter.  Recent 
observation of a medium sized neutron star heavier than twice the solar 
mass~\cite{Margalit:2017dij, Shibata:2017xdx,Rezzolla:2017aly,Ruiz:2017due}, 
simultaneous mass radius measurements~\cite{Riley:2019yda, Miller:2019cac, Riley:2021pdl, 
Miller:2021qha,Fonseca:2021wxt} and bounds on 
tidal deformability~\cite{LIGOScientific:2017vwq, Hinderer:2009ca}
during neutron star mergers have lead to more stringent constraints on the nuclear 
EoS~\cite{Douchin:2001sv, Chatziioannou:2015uea,Annala:2017llu, Tews:2018iwm, Nandi:2018ami, McLerran:2018hbz,Annala:2019puf, Forbes:2019xaz, Baym:2019iky, Drischler:2020fvz, Kojo:2020ztt, Drischler:2020hwi} and hence on these mean-field models.

In this Letter we address the question of how well these nuclear models which 
are traditionally designed to explain cold-dense nuclear matter, can describe QCD 
thermodynamics at a relatively higher temperatures and moderate baryon densities. 
By extending the study of relativistic hadron models (which are well 
constrained from neutron-star merger observations) to moderate values of 
baryon densities and higher temperatures i.e. in the region $\mu_B/T=3$-$5$, 
we estimate the hadronic freezeout line. Using the information about the
pseudo-critical line at small baryon densities from lattice QCD, we constrain
the location of the critical end-point (CEP) at $\mu_{\text{CEP}}/T_{\text{CEP}}>5$ 
if $T_{\text{CEP}}\sim 0.8~T_c$~\cite{Bazavov:2017dus}. The broad outline of the 
Letter is as follows: we begin in the next section by introducing the specific 
relativistic mean-field hadron model used in this work. In the subsequent sections 
we calculate different thermodynamic observables within these mean-field models and 
compare to the lattice QCD results on thermodynamic susceptibilities at finite 
temperatures and moderate values of net-baryon densities.  We show that these mean-field 
models can be very well extended beyond its traditional regime of application and in some 
cases can explain lattice data better than QMHRG. This has deeper implications and 
suggest that non-resonant interactions between the hadrons are crucial to explain 
the lattice QCD data at moderate baryon densities revealing the universal nature of 
hadronic interactions.

\textbf{Relativistic mean-field models for hadrodynamics :}
In order to choose a suitable starting point we consider a 
relativistic mean-field effective model which includes 
strange baryons~\cite{Bunta:2004ej}. This specific model 
along with other two models~\cite{Liu:2001iz, Typel:1999yq} 
are very well constrained out of many hadronic models using the 
latest gravitational wave data coming from neutron star mergers 
and experimental data on nuclear skin thickness~\cite{Nandi:2018ami}.  
Indeed the recent ab-initio result for the 
neutron skin of $\text{Pb}^{208}$~\cite{Hu:2021trw} is consistent with 
this model. The Lagrangian of this effective model~\cite{Bunta:2004ej} 
is explicitly discussed in Appendix I. We henceforth 
refer to this as model 1. In this model, the nucleon fields are coupled non-linearly to the Lorentz scalars 
$\sigma, \delta$ and vectors $\rho, \omega$ mesons. The interactions 
mediated by $\delta$-mesons contribute to the asymmetry energy between 
protons and neutrons and is important for the stability of the nuclei 
drip line. One has additionally the hyperon interactions 
built in it. The strength of hyperon interactions are constrained from different 
sources. Whereas the coupling of hyperons to vector and iso-vector mesons both 
with and without strangeness are  constrained from $SU(6)$ symmetry within the 
quark model, their coupling to scalar mesons are constrained by reproducing the 
hyper-nuclear potentials in saturated nuclear matter. 

At the mean field level, the contribution from the $\omega$-condensate is essentially
proportional to the baryon density and that from the mean $\sigma$ field 
depends on the sum of baryon and anti-baryon densities, see Appendix II. Hence the effect 
of repulsive $\omega$ interactions in thermodynamic observables are visible only at finite 
net-baryon density, within this approximation. We will henceforth show our 
results at finite $\mu_B$ for two cases, i) $n_Q/n_B=0.4,~n_S=0$, the so-called 
strangeness neutral conditions that are realized in a typical heavy-ion collision 
for the phase diagram and ii) $\mu_Q=0, \mu_S=0$ where most lattice QCD data of correlations and fluctuations of baryon number, strangeness, etc are available for comparison. 
In these models the mean field values of meson fields at different temperature and densities 
are obtained by solving a set of self-consistent equations corresponding 
to the nucleon masses and energies, the details of which are discussed in 
Appendix I. The mean fields of the different mesons which mediate  
interactions between baryons in model 1 are shown in Fig.~\ref{fig:meanfields}. 
It is evident that the mean-values of the $\sigma$ field are most 
sensitive to $\mu_B$ followed by the $\omega$ field. Hence for 
$\mu_B<300$ MeV, the attractive interactions due to $\sigma$ dominates 
over the repulsive vector interactions due to $\omega$ mesons. Since the 
strange mesons are massive, their mean-fields have a negligibly small 
dependence on $\mu_B$.

\textbf{How can mean-field models be extended to explain QCD thermodynamics in $T$-$\mu_B$ plane :}

As mentioned earlier, the mean-field models were traditionally introduced to explain 
the nuclear liquid-gas transition in the $T\sim 0$ and large $\mu_B$ regime. We suggest 
here how we can extend the applicability of such models in the finite temperature and 
moderately high baryon number densities.  We first study the ratio 
$\chi^{BS}_{31}/\chi^{BS}_{11}$ where we motivate the need to include additional 
baryons in the spectrum with suitably tuned couplings and in addition non-interacting 
mesons.

\begin{figure}
\includegraphics[scale=0.5]{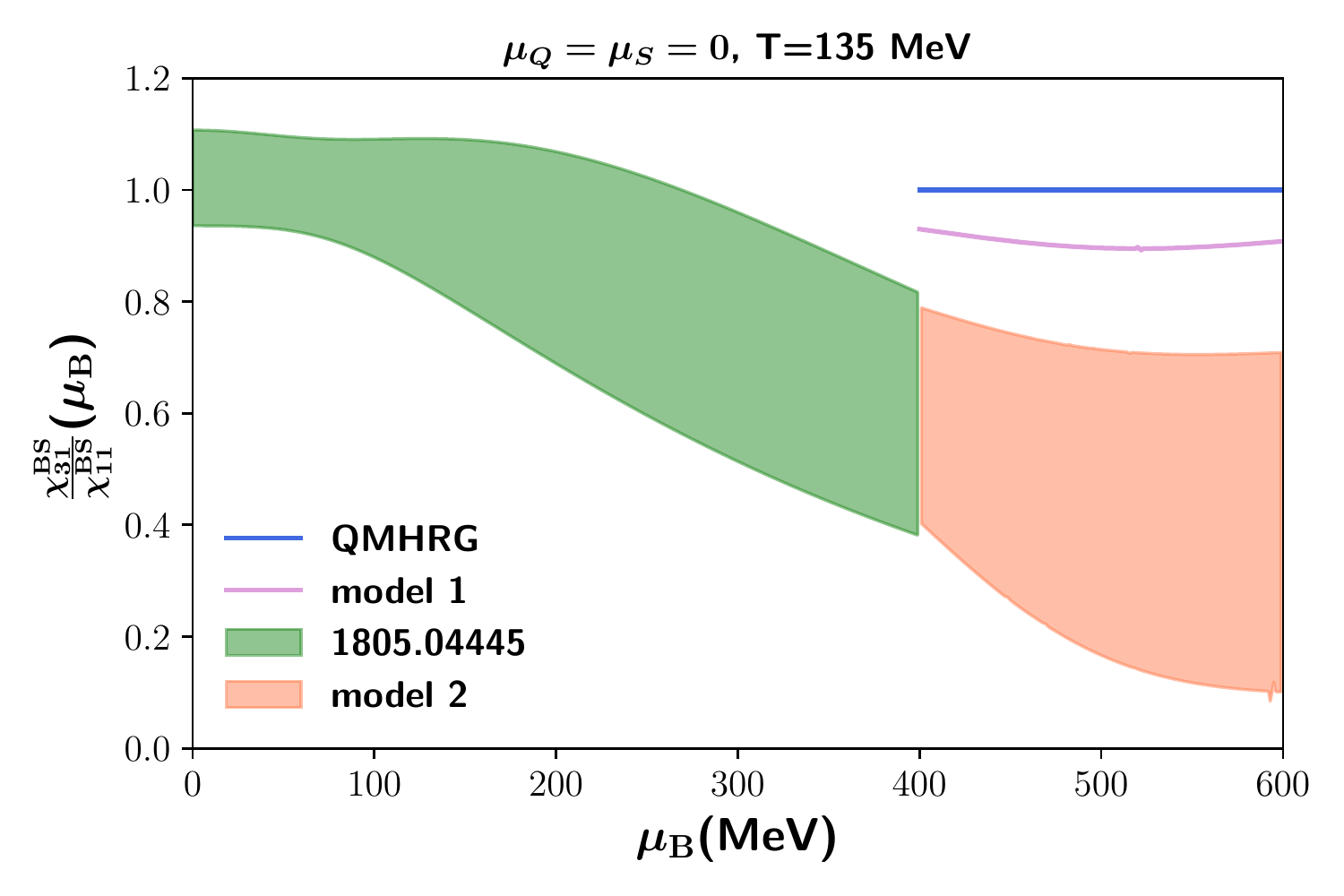}
\caption{The $\chi^{BS}_{31}/\chi^{BS}_{11}$ compared between lattice QCD (green band), QMHRG model (blue), model 1 (magenta) and model 2 (orange band). For model 2 the band corresponds to $\alpha$ varying from $0.15$ to $0.2$ and $\alpha_S$ varying from $0.15$ to $0.7$.}
\label{chi3BS1}
\end{figure}

We show the ratio $\chi^{BS}_{31}/\chi^{BS}_{11}$ as a function of $\mu_B$ at $T=135$ MeV and 
$\mu_Q=\mu_S=0$ in Fig.~\ref{chi3BS1} for $\mu_B>400$ MeV. The lowest value of $\mu_B$ was chosen 
in order to have significant effect of repulsive interactions to this observable while the upper 
limit of $\mu_B$ is chosen such that the energy density at $T=135$ MeV is close to $\epsilon=348
\pm 41$ MeV/$\text{fm}^3$.  The green band represents the results from lattice QCD~\cite{Borsanyi:2018grb} 
which is extrapolated upto $\mu_B=400$ MeV. Results from QMHRG model and  model 1 are also shown in the same 
plot from $\mu_B=400$ MeV to $\mu_B=600$ MeV as solid lines in the right hand corner. 
The results from model 1, are significantly different from QMHRG model results and is $\sim 12\%$ higher 
than the upper band of the lattice QCD result at $\mu_B=400$ MeV. Thus while the inclusion of interactions 
which are inbuilt in model 1 improves the approach towards explaining the lattice data, clearly the 
presence of just two strange baryons is not enough. 
	
 This discrepancy motivates the need for extending the model 1 by including more baryons. 
 Adding more baryons to model 1 requires the knowledge of the couplings of these additional 
 degrees of freedom to the meson fields. These couplings cannot be fixed from experiments as 
 not enough data is available. Another alternative is to use group theoretical arguments to 
 relate the couplings of heavier baryons to those baryons whose couplings are known. To 
 simplify our calculation, we instead fix the couplings of these extra baryons in a way 
 such that couplings of non-strange degrees of freedom are taken to be a fraction 
 of nucleon couplings and strange baryons to be a fraction of $\Lambda$-hyperon couplings. 
 This can be mathematically written in a compact form as $g^{B-M}$=$\alpha g^{N-M}$ and 
 $g^{SB-M}$=$\alpha_S g^{\Lambda-M}$, where $B$ and $SB$ denote non-strange and strange baryons 
 respectively which are not present in model 1 but included from the QMHRG list. The $N$ represents 
 nucleon and $M$ denotes mesons. The coefficients  $\alpha$ and $\alpha_S$ are obtained by fitting model 2 
 results of $\chi^{BS}_{31}/\chi^{BS}_{11}$ to the continuum extrapolated band from lattice QCD. The
 model 2 results  match with the upper boundary of lattice QCD data for $\alpha = \alpha_S = 0.15$. 
 To also match the model 2 results to the lower boundary of the lattice QCD data,  we have to choose 
 $\alpha = 0.2$ and $\alpha_S = 0.7$. Thus between the upper and lower edges of the lattice 
 data in Fig.~\ref{chi3BS1}, the non-strange baryon couplings change little while the strange 
 baryon couplings vary significantly.

\begin{figure}
\includegraphics[scale=0.5]{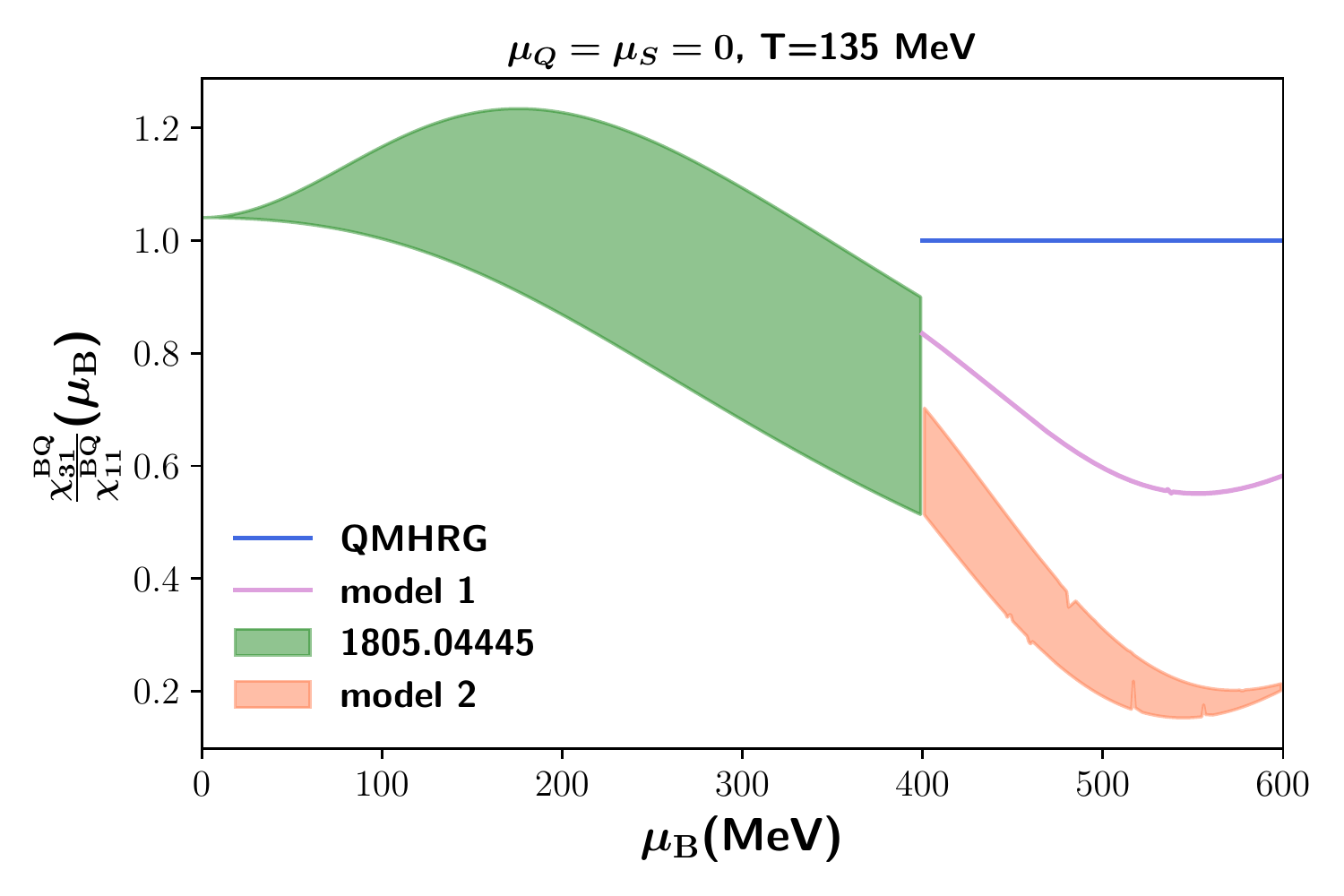}
\caption{The $\chi^{BQ}_{31}/\chi^{BQ}_{11}$ compared between lattice QCD (green band), QMHRG model (blue), model 1 (magenta) and model 2 (orange band).}
\label{chi3BQ1}
\end{figure}

Now having fixed the couplings, we show what are its implications for other thermodynamic 
observables. In Fig.~\ref{chi3BQ1} we show the results for $\chi^{BQ}_{31}/\chi^{BQ}_{11}$ 
within model 2. Comparing with the lattice QCD results we find that the results 
from model 1 already agree with the lattice band but towards its upper edge. Now calculating the 
same observable in the model 2 with additional hadrons, we find an agreement with the lattice QCD 
results within a more constrained region with the lower boundaries of these bands agreeing well at 
$\mu_B=400$ MeV. We recall here that the spread in model 2 results comes from the variation of $\alpha$ 
from $0.15$ to $0.2$ and $\alpha_S$ from $0.15$ to $0.7$.  

\begin{figure}
\includegraphics[scale=0.5]{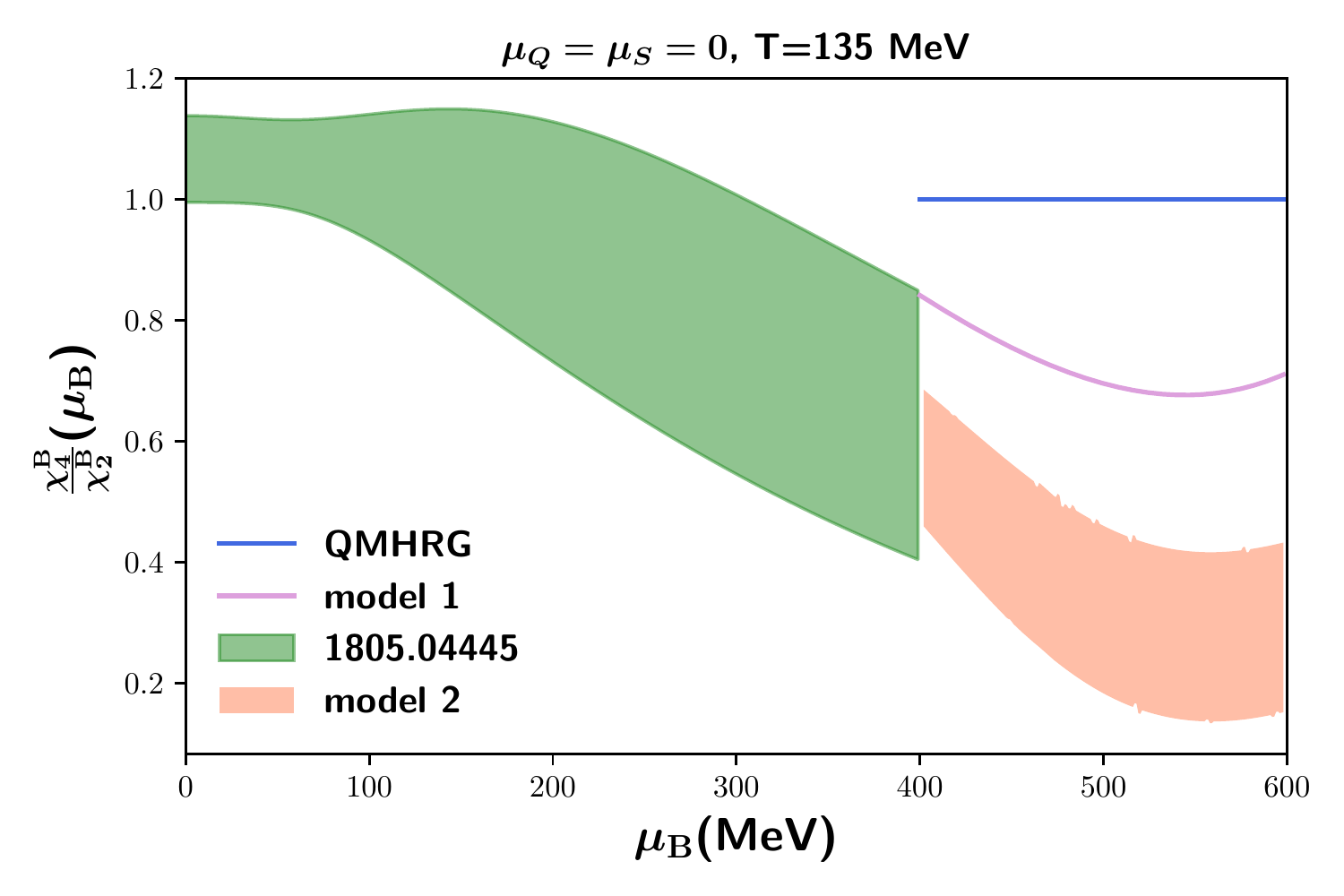}
\caption{The $\chi^{B}_{4}/\chi^{B}_{2}$ compared between lattice QCD (green band), QMHRG model (blue), model 1 (magenta) and model 2 (orange band).}
\label{chi4chi2}
\end{figure}

We next show the results of another interesting observable $\chi^{B}_{4}/\chi^{B}_{2}$ as a function of 
$\mu_B$ at $T=135$ MeV and $\mu_Q=\mu_S=0$. Again, the calculations within the model 1 
agrees with the upper boundary of lattice band. Using model 2, we find 
that the results of this ratio has a smaller spread which arise due to the uncertainty in the 
values of the heavier baryon-meson couplings, compared to the current error band in the lattice QCD 
data. 

\textbf{Do extended mean-field models satisfy high density constraints?}

\begin{figure}
\includegraphics[scale=0.5]{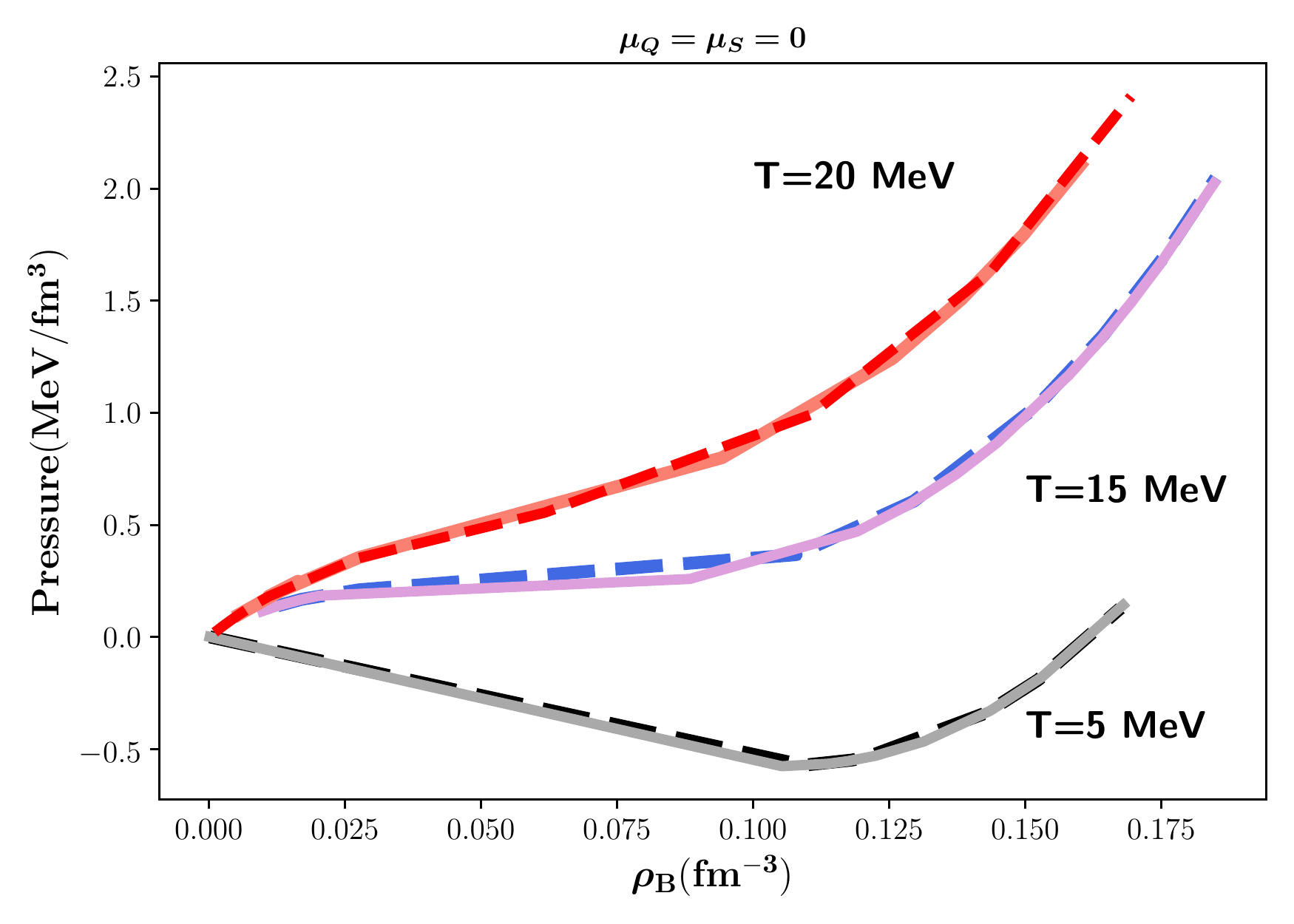}
\caption{Pressure as a function of baryon number density calculated from model 1 (dashed lines) and 2 (solid lines) for three different temperatures. The $ T \sim 15$ MeV corresponds to nuclear liquid gas transition.}
\label{fig:constraint1}
\end{figure}

Since the model 1 satisfies very well the constraints from high density matter, like nuclear liquid-gas 
transition, neutron star EoS, etc,  we will check whether augmenting this model with additional hadrons 
would in anyway affect this agreement. While the contribution of heavier baryons at high densities and 
low temperatures are expected to be suppressed due to the thermodynamic distribution functions, their 
multiplicities are large which could influence the thermodynamics despite suppression of thermal distributions. 
To check how much of an impact these additional baryons will make, we calculate the pressure as a function of 
baryon densities within model 1 and model 2 for three different temperatures $T=5, 15, 20$ MeV respectively. 
The ratios of couplings used for model 2 calculations have been chosen to be $\alpha=0.2$ and $\alpha_S=0.7$.  
The results of our calculation are shown in Fig.~\ref{fig:constraint1} as solid lines for model 2 which is 
compared with the dashed lines in the same plot for the  model 1. 
As clearly seen from the plot, extending the model with the additional hadrons from QMHRG model with 
a large spread in the allowed values of couplings does not affect the pressure vs density curves.  
This gives us a proof of principle that extending mean-field nuclear models with additional baryons to 
explain QCD thermodynamics at high temperatures and intermediate densities will not affect its already 
excellent agreement at high densities and low temperatures. Our approach indeed hints to a method towards 
formulating a universal theory describing the hadronic phase of QCD.

\textbf{Implications for the phase diagram of 
QCD :}
\begin{figure}[h]
\includegraphics[scale=0.5]{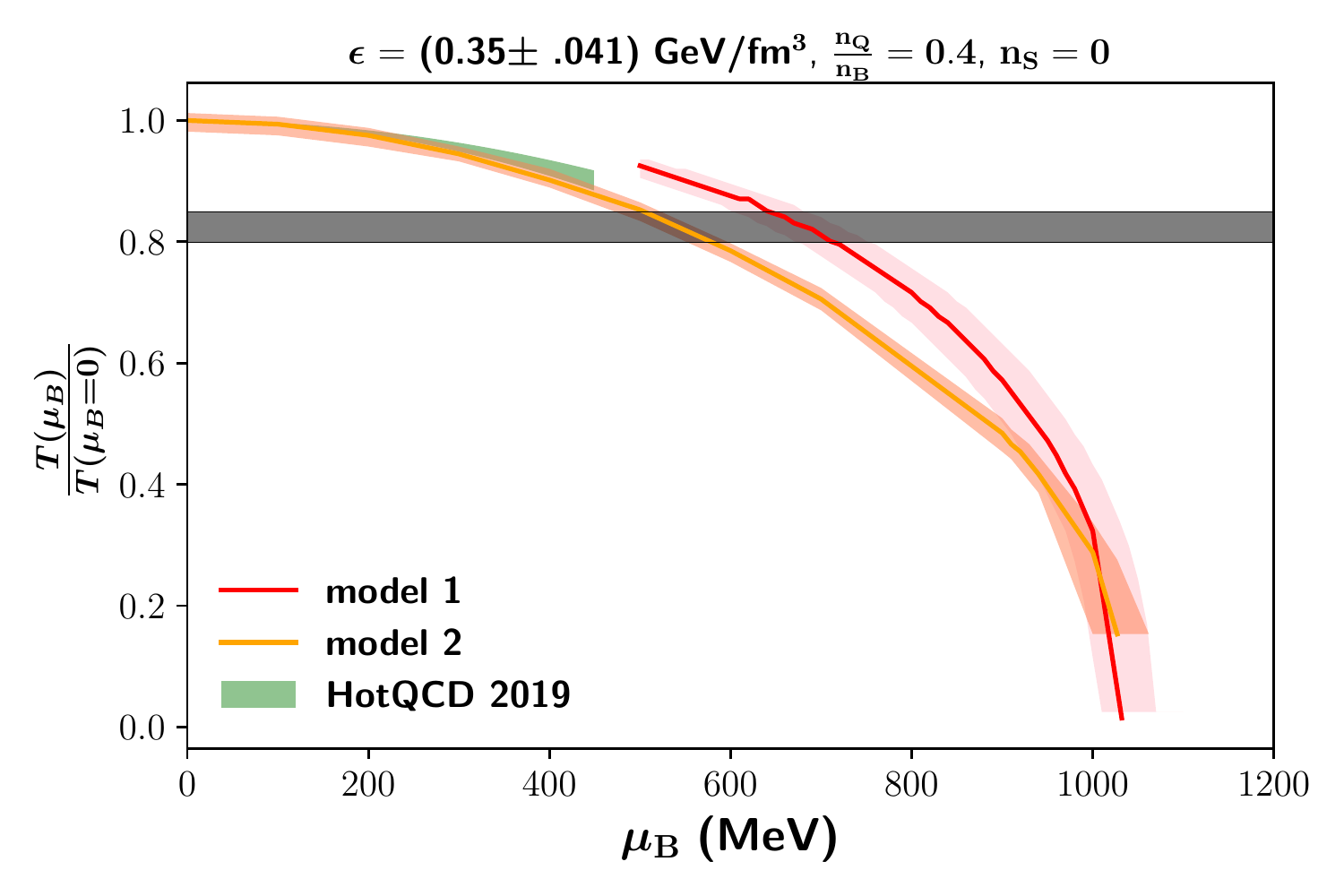}
\caption{Lines of constant energy density for model 1 shown as a red band and model 2 shown as an orange band. These results are compared with the chiral crossover line obtained from lattice QCD~\cite{Bazavov:2017dus} and shown as a green band.}
\label{fig:constantELine}
\end{figure}

Having discussed the susceptibilities in the mean-field nuclear model and its extended version, we 
study what insights it can give us about the phase diagram of QCD. Nuclear mean-field models do 
not have the $U_L(2)\times U_R(2)$ chiral symmetries in-built like the Nambu-Jona-Lasinio model 
and hence cannot describe its restoration. We thus determine the line of constant energy density in the 
$T$-$\mu_B$ plane for these models by setting $\epsilon= 348 \pm 41$  MeV/$\text{fm}^3$~\cite{Bazavov:2017dus} 
which is the energy density of $2+1$ flavor QCD at the crossover region for $\mu_B= 0$ MeV. Unlike in 
traditional QMHRG model, recent lattice studies have observed that the energy density along the chiral 
crossover line does not vary with increasing $\mu_B$  at least around $\mu_B/T\lesssim 3$~\cite{Bollweg:2022fqq}. 
Incidentally the line of chemical freezeout of hadrons is also defined at a constant energy 
density~\cite{Becattini:1997ii,Teaney:2002aj} and it approaches the chiral crossover 
transition line as one goes to smaller values of $\mu_B$. The results of our 
calculations of lines of constant energy density in model 1 and 2 are shown in 
Fig.~\ref{fig:constantELine}. These can be visualized as a chemical freezeout lines for 
the hadrons present within the model. Indeed the line of constant energy density in model 2
is consistent with the latest continuum extrapolated lattice QCD data,  all the way from 
$\mu_B = 0$ MeV (extrapolated) to about $\mu_B = 450$ MeV. There is a small difference between 
these two results which can be accounted for from the fact that the repulsive interactions present in model 1 and 2 are insignificant at lower values of $\mu_B$.  Since the model 2 has more degrees 
of freedom, its line of constant energy density deviates from the model 1 calculation for 
$\mu_B<900$ MeV. At higher values of $\mu_B$, the contributions of the heavier baryons and mesons 
to the energy density gets suppressed due to their mass and due to lowering of temperature respectively, 
hence the lines of constant energy between model 1 and its extended version start to agree.  Another 
prominent feature of the QCD phase diagram is the anticipated critical end-point (CEP) of the line 
of first order transitions.  From the constraint that the CEP will exist in the real-$\mu_B$ plane, and its
location gives the radius of convergence of thermodynamic observables, all orders of baryon number 
fluctuations have to be positive. Using this constraint from the lattice QCD data upto $8$-th order 
baryon number fluctuations at $\mu_B=0$~\cite{Bazavov:2017dus},  it is now known that $T_{\text{CEP}}/T_c<0.85$. 
Noting this constraint by choosing the ratio  $T/T(\mu_B=0) = 0.8$ within the model 2 we can conclude that 
the CEP, if present will be at $\mu_B > 596$ MeV which provides a lower bound $\mu_B/T \sim 4.76$.

Next we calculate the curvature of these constant energy lines by fitting to the ansatz 
$\frac{T(\mu_B)}{T_c}=1-\kappa_2 \frac{\mu_B^2}{T_c^2}-\kappa_4 \frac{\mu_B^4}{T_c^4}-\kappa_6 \frac{\mu_B^6}{T_c^6}$. For model 1, the extracted curvature coefficients are $\kappa_2=0.020(2)$, 
$\kappa_4=-0.0010(3)$, $\kappa_6=0.000060(3)$ which are also consistent with those 
calculated from model 2, $\kappa_2=0.020(2)$, $\kappa_4=-0.0005(1)$ and 
$\kappa_6=0.000010(2)$. The values of $\kappa_2$ are somewhat larger than the latest 
continuum extrapolated lattice results of the $\kappa_2$~\cite{Bazavov:2017dus,Bonati:2018nut,Borsanyi:2020fev} 
extracted from the renormalized chiral condensate and from a recent HRG model estimate~\cite{Biswas:2022vat}, 
which is expected as the results from these models are for the entire $T$-$\mu_B$ plane.  The value 
of $\kappa_4$ from lattice QCD is consistent with zero~~\cite{Bazavov:2017dus,Borsanyi:2020fev}, 
whereas we find a negative but finite value in both the models. The results for $\kappa_6$ are new 
and it is about $1000$ times smaller than $\kappa_2$. Thus its effect should start become significant 
at $\mu_B/T\sim 15$, well within the cold nuclear matter regime. Moreover the baryon densities obtained 
in model 1 and 2 for a typical neutron star environment characterized with $n_Q/n_B=0.05$-$0.2, n_S=0$, 
varies from $0.28$ fm$^{-3}$~ to $0.35$ fm$^{-3}$ as energy density varies from  
$\epsilon= 307$-$389$ MeV/$\text{fm}^3$. The variation in the ratio for $n_Q/n_B$ has
a tiny effect on this density. It is remarkable that the typical nuclear densities we 
obtain from these models are about twice the nuclear saturation density, when many-body 
interactions start to become dominant~\cite{Akmal:1998cf} and 
quark exchanges are expected to mediate baryon interactions~\cite{Fukushima:2020cmk}. 
Our calculations also support this picture albeit indirectly  that a mixed phase of quarks 
and hadrons can survive in neutron star cores with baryon densities greater than $0.35$ fm$^{-3}$.

\textbf{Implications of lattice QCD data at $\mu_B=0$ for high density models :}
Comparisons of lattice QCD data with QMHRG model particularly for observables like $\chi^{B}_{4}/\chi^{B}_{2}$ 
~\cite{Bollweg:2021vqf} and higher order baryon number susceptibilities~\cite{Huovinen:2009yb} 
clearly highlight the importance of including repulsive interactions within the QMHRG model. 
In our present study of nuclear model quantum field theories, the repulsive interactions at 
low baryon densities are negligible at the mean-field level. Unless there is a mechanism by which 
sufficient strength of repulsive interactions are generated at low baryon densities 
by calculating beyond mean-field effects, it would then imply that these models require suitable modifications 
to account for such interactions. In this way one can achieve a universal hadronic model, which is valid 
for both lower as well as high baryon densities.  Furthermore our comparison of quantities like $\chi^{BS}_{31}/
\chi^{BS}_{11}$ with the lattice data to extract the baryon-meson couplings in the  model 2, will benefit 
from an increasing precision of the lattice QCD data. This will allow for a tighter constraint on the values 
of the couplings of strange baryons with mesons.

\textbf{Conclusions :} 
We started this work with a question of how well the traditional 
nuclear mean-field models, developed for the understanding of physics 
at low temperatures and large baryon densities can be used to explain QCD 
thermodynamics at high temperatures and moderate densities. A remarkable 
observation that comes out of our study is that augmenting these simple 
models with a complete list of baryons present in QMHRG model and tuning 
the couplings of their interactions with mesons through a comparison with 
lattice QCD data of a particular observable, leads to a very good description 
of QCD thermodynamics at intermediate densities. In our investigation we have 
found that the simple baryon-meson interactions built within the nuclear models 
are important in bridging the gap between lattice and other non-interacting hadron 
models like QMHRG.  Furthermore we have shown that the inclusion of these additional 
hadrons do not affect the nuclear liquid-gas transition, which is well-studied in the 
original versions of these mean-field models.

This allows for a route to identify the relevant baryon interactions in chiral 
symmetry broken phase, which indeed if accounted for correctly will be valid 
for the entire regime of densities and temperatures. However at 
present there are not much data available, either from experiments or theory in 
constraining most of these baryon-meson couplings.  Our method for determining 
these couplings from comparison with a particular thermodynamic observable from lattice, 
is one such possibility since in this process the benchmark data comes from the fundamental 
theory of strong interactions i.e., QCD. We highlighted the need of high-precision lattice 
data which will allow for constraining such couplings further. This will allow for a 
better synergy between lattice QCD and such model quantum field theory calculations in future. 

There are several directions still remaining  to be explored. Firstly it would be interesting 
to extend this study beyond the mean-field approximation and check whether it can 
account for the repulsive interactions that exist among baryons and mesons, even at low densities 
and high temperatures, evident from comparisons of lattice QCD data with QMHRG. Another aspect 
towards building a universal hadronic model requires high density nuclear models to incorporate 
spontaneous chiral symmetry breaking. This can be achieved by including parity doublet partners like 
the pion degrees of freedom and the critical $\sigma$-modes, important for understanding the nature of 
the chiral phase transition at high densities and the thermodynamics near the critical end-point.

\textbf{Acknowledgements :}
S.S. gratefully acknowledges support from the Department of Science and Technology, 
Government of India through a Ramanujan Fellowship. We would like to thank Deeptak 
Biswas, Jishnu Goswami, Hiranmaya Mishra and Jan Pawlowski for helpful discussions and 
correspondence.

\begin{appendix}
\section{Appendix I: Details of the numerical calculations with Model 1}
\label{sec:app1}
The Lagrangian describing model 1~\cite{Bunta:2004ej} is given as :
\begin{equation}
\begin{split}
&\mathcal{L} = \bar{\psi}\big[\gamma_\mu (i \partial^\mu - g_\omega \omega^\mu - g_\rho \rho^\mu  \tau -g_{\phi B} \phi^\mu) \nonumber \\
&- (M - g_\sigma \sigma -g_\delta \delta \tau - g_{\sigma^* B}\sigma^*)\big]\psi  \nonumber \\
& +\frac{1}{2} (\partial_\mu \sigma \partial^\mu \sigma - m_\sigma^2 \sigma^2) -\frac{1}{3}b_\sigma M (g_\sigma \sigma)^3 - \frac{1}{4} c_\sigma (g_\sigma \sigma)^4  
  \nonumber \\
& -\frac{1}{4} (\omega_{\mu\nu} \omega^{\mu\nu}) +\frac{1}{2} m_\omega^2 (\omega_\mu \omega^\mu) + \frac{1}{4}c_\omega (g_\omega^2 \omega_\mu \omega^\mu)^2 \nonumber \\
& +\frac{1}{2} (\partial_\mu \delta \partial^\mu \delta - m_\delta^2 \delta^2) + \frac{1}{2} m_\rho^2 \rho^\mu . \rho_\mu - \frac{1}{4} (\rho_{\mu\nu} \rho^{\mu\nu}) \nonumber \\
& +\frac{1}{2} \Lambda_V (g_\rho^2 \rho_\mu . \rho^\mu) (g_\omega^2 \omega_\mu \omega^\mu)
 + \frac{1}{2} (\partial_\mu \sigma^* \partial^\mu \sigma^* - m_\sigma^{*2} \sigma^{*2}) \nonumber \\ 
&+ \frac{1}{2} m_\phi^2 \phi_\mu \phi^\mu - \frac{1}{4} \phi_{\mu\nu} \phi^{\mu\nu} \nonumber \\
& + \sum_{e,\mu} \bar{\psi}_{e,\mu} (i \gamma_\mu \partial^\mu - m_{e,\mu})\psi_{e,\mu} .
\end{split}
\end{equation}

Using this Lagrangian we have derived the equation of motion for the meson fields in the mean-field 
approximation. In this approximation the meson fields are approximated by the spacetime independent 
values satisfying the equations of motion and constraints for the net-electric charge and strangeness 
densities. The mean field equations which we solve are,

\begin{equation*}
m_\sigma^2 \sigma = g_\sigma \bigg[ \sum_B \frac{g_{\sigma B}}{g_\sigma} \rho_B^S -b_\sigma M (g_\sigma \sigma)^2 - c_\sigma (g_\sigma)^3  \bigg]
\end{equation*}
    
\begin{equation*}
m_\omega^2 \omega^\nu = g_\omega \bigg[ \sum_B \frac{g_{\omega B}}{g_\omega} \rho_B^B -c_\omega M (\omega_\mu \omega^\mu \omega_\mu ) - g_\rho^2 \rho_\mu . \rho^\mu \Lambda_V g_\omega \omega_\mu  \bigg]
\end{equation*}

\begin{equation*}
m_\rho^2 \rho^\nu = g_\rho \bigg[ \sum_B \frac{g_{\rho B}}{g_\rho} \rho_B^B \tau - g_\rho \rho_\mu \Lambda_V g_\omega^2 \omega_\mu \omega^\mu  \bigg]
\end{equation*}

\begin{equation*}
m_\delta^2 \delta = g_\delta \sum_B \frac{g_{\delta B}}{g_\delta} \rho_B^S \tau
\end{equation*}

\begin{equation*}
m_{\sigma^*}^2 \delta = g_{\sigma^* \Lambda} \sum_B \frac{g_{\sigma^* 
 B}}{g_{\sigma^* \Lambda}} \rho_B^S 
\end{equation*}

\begin{equation*}
m_\phi^2 \phi^\nu = g_{\phi \Lambda} \sum_B \frac{g_{\phi 
 B}}{g_{\phi \Lambda}} \rho_B^B .
\end{equation*}

The values of the different couplings and other details can be found in Ref.~\cite{Bunta:2004ej}. 
Once the values of mean fields are known from the solutions of these equations satisfying the constraints, 
the pressure can be calculated. The susceptibilities can then be calculated by taking the derivatives 
of pressure with respect to different chemical potentials corresponding to baryon number, strangeness, etc. 
We schematically show our calculation in the following equations where $X_i$ are the mean fields, $\mu_j$ denote 
the chemical potentials and $\vec{f}$ denote the gap equations obtained from equations of 
motion and $\vec{g}$ are the constraints on the system. 

\begin{eqnarray}
        P(X_i,\mu_j,T) = 0 ,\\
	     \vec{f}(X_i,\mu_j,T) = 0 ,\\
        \vec{g}(X_i,\mu_j,T) = 0 .
\end{eqnarray}

To calculate the derivatives one can use the finite difference method which is numerically 
accurate upto $\mathcal{O}(\delta \mu_k)$. Moreover the truncation error increases with increasing 
order of the derivatives. We thus follow a different procedure. Instead of differentiating 
numerically, we use the gap and constraint equations to calculate the derivatives analytically. This is 
possible because these equations are satisfied at each values of $T$ and $\mu_k$, and the total derivative 
of each of them is zero.

\begin{eqnarray}
  \frac{\partial \vec{f}}{\partial \mu_k} + \frac{\partial \vec{f}}{\partial X_i} \frac{dX_i}{d\mu_k} + \frac{\partial \vec{f}}{\partial \mu_j} \frac{d\mu_j}{d \mu_k} = 0 ,\\
  \frac{\partial \vec{g}}{\partial \mu_k} + \frac{\partial \vec{g}}{\partial X_i} \frac{dX_i}{d\mu_k} + \frac{\partial \vec{g}}{\partial \mu_j} \frac{d\mu_j}{d\mu_k} = 0 .
\end{eqnarray}

 These are linear equations in the derivatives $\frac{dX}{d\mu_k}$ and $\frac{d\mu_j}{d\mu_k}$. 
 Solving for the above equations gives the first order derivatives of mean fields. Once first order 
 derivatives are known, these equations can be successively differentiated to find further 
 higher order derivatives.

\section{Appendix II : The various meanfields as a function of $\mu_B$}
\label{sec:app2}
\begin{figure}
\includegraphics[scale=0.5]{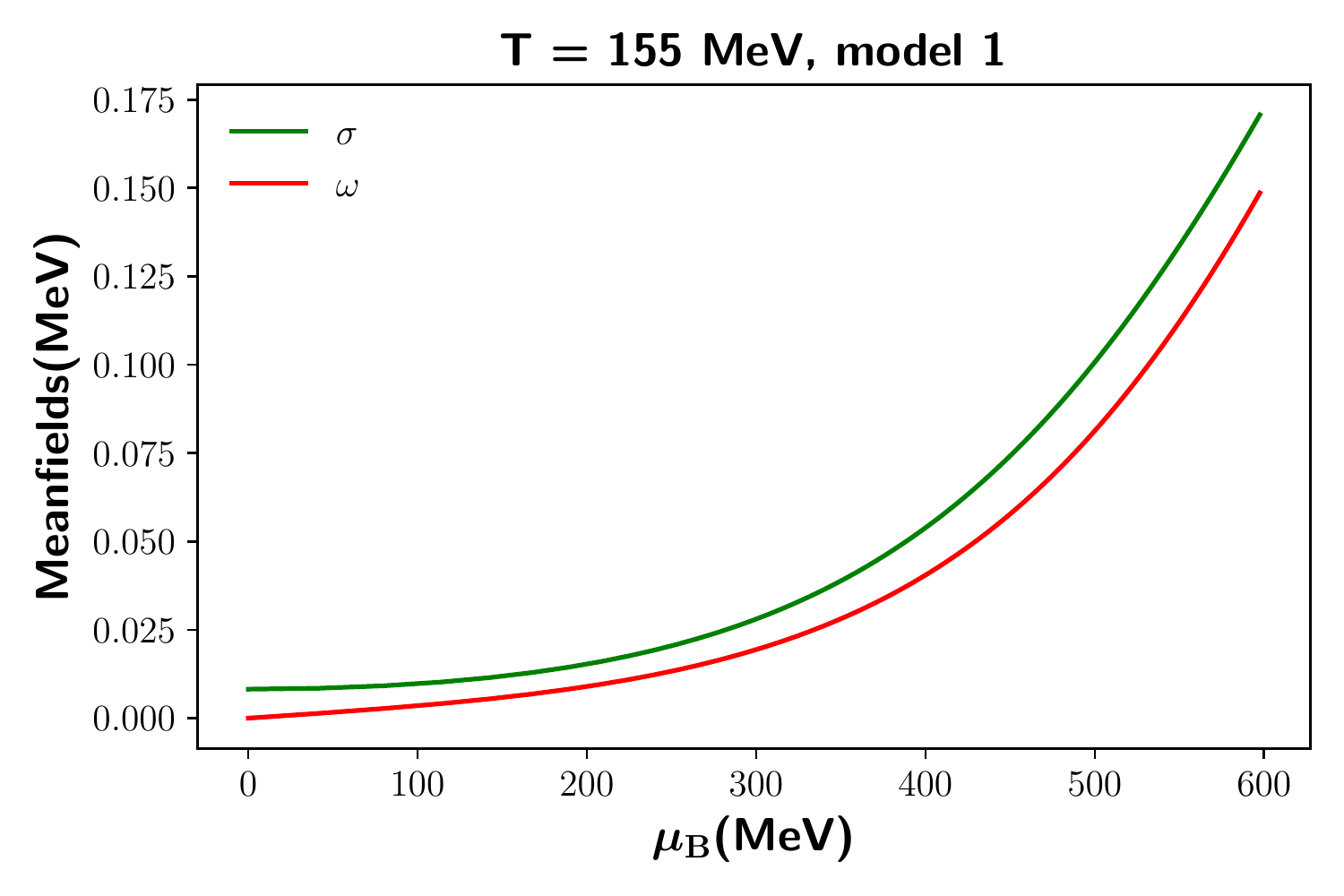}
\caption{The $\sigma$ and $\omega$ mean-field value as a function of $\mu_B$ for $T=155$ MeV. Other mean-fields are much smaller.}
\label{fig:meanfields}
\end{figure}

In Fig.~\ref{fig:meanfields} we have shown $\sigma$ and $\omega$ mean fields in model 1~\cite{Bunta:2004ej} as a function of 
the baryon chemical potential. From the plot we observe that as $\mu_B$ increases, the mean-field values of 
$\sigma$ and $\omega$ fields also increases, hence the interactions mediated by these mesons become more 
relevant. This is because the mean-field values for $\sigma$ and $\omega$ fields are proportional to scalar and 
baryon densities respectively, which increase with the baryon chemical potential. The $\rho$ and $\delta$ mean-fields 
are proportional to the isospin baryon and scalar densities respectively. Since the isospin chemical potential is 
negligibly small, these mean-fields remain insignificant. The $\sigma_s$ and $\phi_s$ mesons couple only to strange 
baryons which are heavy and thus their mean-field values remain small. The mean-fields which increase the pressure 
correspond to mesons which mediate repulsive interactions and which decreases the value of pressure correspond to 
attractive interactions. The $\omega$ thus mediates repulsive interactions and the $\sigma$ mediates attractive 
interactions. One may note that while fields other than $\sigma$ and $\omega$ are small and not shown in  
Fig.~\ref{fig:meanfields}, they play an important role at high density and low temperature in fitting 
the experimental data.

\end{appendix}

\bibliography{ref}
\bibliographystyle{ieeetr}	

\end{document}